# On Lorentz and Einstein-Laub forces in material media

A L Kholmetskii[1], O V Missevitch[2], T Yarman[3,4]

[1]Belarus State University, 4 Nezavisimosti Avenue, 220030 Minsk, Belarus
[2]Institute for Nuclear Problems, Belarus State University, 11 Bobruiskaya Street, 220030 Minsk, Belarus
[3]Okan University, Akfirat, Istanbul, Turkey
[4]Savronik, Eskisehir, Turkey

As a continuation of the discussion started in (M. Mansuripur, Phys. Rev. Lett. **108**, 193901 (2012)), we show that the approach based on Lorentz force law in material media, like Einstein-Laub expression for electromagnetic force, gives relativistically consistent results, when the contribution of hidden momentum is properly accounted for. An expression for the torque on point-like electric/magnetic dipole is derived, which is compatible with both the Lorentz and Einstein-Laub approaches.

In a recent letter [1] Mansuripur questioned the correctness of Lorentz force law in material media, arguing that the force law by Einstein and Laub [2, 3] must be adopted instead, insofar as only the latter law provides the implementation of momentum conservation law and exhibits compatibility with special relativity. As a central point of his argumentation, the author of [1] considers the interaction between a point-like charge $q$ and point-like magnetic dipole $\boldsymbol{\mu}$, which are both at rest with respect to each other in a reference frame K', and the direction of $\boldsymbol{\mu}$ is orthogonal to the line joining charge and dipole (the axis $z$). In this frame the mutual force between charge and magnetic dipole is equal to zero, and no torque is exerted on the dipole by the resting charge. Then he considers the situation, where the frame K' is moving at the constant velocity $V$ along the axis $z$ of the laboratory frame K, and shows that in this frame the forces acting on charge and dipole remain to be zero. However, the Lorentz force law

$$\boldsymbol{f}_L = \rho_{total}\boldsymbol{E} + \boldsymbol{J}_{total}\times\boldsymbol{B} \tag{1}$$

(along with the designations used in ref. [1]) yields a non-vanishing torque exerted on the moving magnetic dipole by the moving charge, which obviously represents a non-adequate result, since in the proper frame of charge and dipole K', the torque is equal to zero. On the other hand, when the Einstein-Laub formula is applied, both the force and the torque are equal to zero in the frames K' and K. Based on this result, Mansuripur concluded that for material media the Lorentz force law must be abandoned in favor of the Einstein-Laub law.

In this respect we notice that the incompleteness of Lorentz force law (1) for material media is, in general, the known fact, if the force contribution caused by the time variation of hidden momentum is not included (see, e.g. [4-10]). This contribution gives the force component

$$\boldsymbol{F}_h = -\varepsilon_0 \frac{d}{dt}(\boldsymbol{\mu}\times\boldsymbol{E}), \tag{2}$$

which is required for the system "material medium plus electromagnetic field" to maintain the balance of momenta, as for the first time was pointed out in ref. [4]. Later various models of magnetic dipoles were involved (see, e.g. [5-9]), in order to explain the origin of force (2), which, in general, may include a non-electromagnetic contribution and thus cannot be accounted directly by the Lorentz law.

At the same time, to our recollection, the problem of determination of torque on a magnetic dipole caused by the hidden momentum contribution is not explicitly solved to the moment. One should notice that even in that cases, when the force due to time variation of hidden momentum (2) is vanishing (like in the problem of ref. [1] about the interaction of charge and magnetic dipole), the related torque, in general, might be a non-zero value. It seems that this possibility was analyzed in ref. [1].

A torque experienced by the magnetic dipole in an electric field due to hidden momentum contribution can be defined by the equation

$$\boldsymbol{T}_h=\int_{\xi}(\boldsymbol{r}\times\boldsymbol{f}_h)d\xi, \tag{3}$$

where $\boldsymbol{f}_h$ is the force density of hidden momentum contribution, $\boldsymbol{r}$ is for the position vector of any point inside the dipole and $\xi$ stands for the volume of dipole. Based on eq. (2), we can conjecture the force density due to variation of hidden momentum in the form:

$$\boldsymbol{f}_h=-\frac{\partial}{\partial t}(\boldsymbol{M}\times\varepsilon_0\boldsymbol{E}), \tag{4}$$

where $\boldsymbol{M}$ is the magnetization of medium. The use of partial time derivative in eq. (4) instead of total time derivative in eq. (2) is related to the fact (usually not pointed out in the literature) that the measurement of magnetization (polarization) is carried out at a point *fixed* in a laboratory, whereas the measurement of magnetic (electric) dipole moment of a moving particle is carried out for a volume *co-moving* to this particle. Anyway, using the operator equality $\frac{\partial}{\partial t}=\frac{d}{dt}-(\boldsymbol{V}\cdot\nabla)$ (where $\boldsymbol{V}$ stands for velocity), and taking into account that the volume integral $\int_{\xi}(\boldsymbol{V}\cdot\nabla)(\boldsymbol{M}\times\boldsymbol{E})d\xi$ via the Gauss theorem can be transformed onto a surface integral (where $\boldsymbol{M}$ disappers), we obtain that

$$\int_{\xi}\boldsymbol{f}_h d\xi=-\varepsilon_0\frac{d}{dt}(\boldsymbol{\mu}\times\boldsymbol{E})=\boldsymbol{F}_h.$$

Thus, the force contribution (4) should be added to the Lorentz force density (1), so that in the absence of free charges the total force density reads:

$$\begin{aligned}\boldsymbol{f}_{total}=\boldsymbol{f}_L+\boldsymbol{f}_h=\rho_{total}\boldsymbol{E}+\boldsymbol{J}_{total}\times\boldsymbol{B}-\frac{\partial}{\partial t}(\boldsymbol{M}\times\varepsilon_0\boldsymbol{E})=\\ -(\nabla\cdot\boldsymbol{P})\boldsymbol{E}+(\nabla\times\boldsymbol{M})\times\boldsymbol{B}+\frac{\partial\boldsymbol{P}}{\partial t}\times\boldsymbol{B}-\frac{\partial}{\partial t}(\boldsymbol{M}\times\varepsilon_0\boldsymbol{E}).\end{aligned} \tag{5}$$

In the letter equation we have used the known relationships for bound charge and current densities (see e.g. [10]): $\rho=-\nabla\cdot\boldsymbol{P}$, $J=\frac{\partial\boldsymbol{P}}{\partial t}+\nabla\times\boldsymbol{M}$ (where $\boldsymbol{P}$ being the polarization).

Therefore, the total torque exerted on an electric/magnetic dipole in an external electromagnetic field is equal to

$$\boldsymbol{T}_{total}=\int_{\xi}d\xi\left[\boldsymbol{r}\times\left(-(\nabla\cdot\boldsymbol{P})\boldsymbol{E}+(\nabla\times\boldsymbol{M})\times\boldsymbol{B}+\mu_0\frac{\partial\boldsymbol{P}}{\partial t}\times\boldsymbol{B}-\frac{\partial}{\partial t}(\boldsymbol{M}\times\varepsilon_0\boldsymbol{E})\right)\right]. \tag{6}$$

Based on eq. (6), let us derive the expression for the torque exerted on a point-like electric/magnetic dipole, moving at the velocity $\boldsymbol{V}$ in the external electric $\boldsymbol{E}$ and magnetic $\boldsymbol{B}$ fields of a laboratory frame. In such derivation we assume that the electric $\boldsymbol{p}=\int_{\xi}\boldsymbol{P}d\xi$ and magnetic $\boldsymbol{\mu}=\int_{\xi}\boldsymbol{M}d\xi$ dipole moments of such dipole are the constant values in its rest frame and adopt that the spatial variation of fields $\boldsymbol{E}$, $\boldsymbol{B}$ at the location of point-like dipole is negligible. With these limitations we obtain for the first term in integrand of eq. (6) (contribution of the Coulomb interation):

$$-\int_{\xi}\boldsymbol{r}\times(\nabla\cdot\boldsymbol{P})\boldsymbol{E}d\xi=\boldsymbol{p}\times\boldsymbol{E}, \tag{7}$$

which can be proven in components. For example for the $z$-component we have:

$$-E_y\int_\xi x\left(\frac{\partial P_x}{\partial x}+\frac{\partial P_y}{\partial y}+\frac{\partial P_z}{\partial z}\right)d\xi+E_x\int_\xi y\left(\frac{\partial P_x}{\partial x}+\frac{\partial P_y}{\partial y}+\frac{\partial P_z}{\partial z}\right)d\xi=$$

$$-E_y\int_\xi\left(\frac{\partial(xP_x)}{\partial x}-P_x+\frac{\partial(xP_y)}{\partial y}+\frac{\partial(xP_z)}{\partial z}\right)d\xi+E_x\int_\xi\left(\frac{\partial(yP_x)}{\partial x}+\frac{\partial(yP_y)}{\partial y}-P_y+\frac{\partial(yP_z)}{\partial z}\right)d\xi=$$

$$p_xE_y-p_yE_x=(\boldsymbol{p}\times\boldsymbol{E})_z$$

Here we have taken into account that the integration of the terms $\partial(r_iP_j)/\partial r_l$ ($i$, $j$, $l$=1…3) over the volume of the dipole gives the value of $P_j$ on its surface, which is equal to zero.

Next we evaluate the contribution due to interaction of magnetization currents of a dipole with a magnetic field (second term in the integrand of eq. (6)). Using the vector identities $\boldsymbol{a}\times\boldsymbol{b}=-\boldsymbol{b}\times\boldsymbol{a}$ and $\boldsymbol{a}\times(\nabla\times\boldsymbol{b})=\nabla(\boldsymbol{a}\cdot\boldsymbol{b})-\boldsymbol{b}\times(\nabla\times\boldsymbol{a})-(\boldsymbol{a}\cdot\nabla)\boldsymbol{b}-(\boldsymbol{b}\cdot\nabla)\boldsymbol{a}$, this term can be presented in the form:

$$\int_\xi\boldsymbol{r}\times((\nabla\times\boldsymbol{M})\times\boldsymbol{B})d\xi=-\int_\xi\boldsymbol{r}\times(\boldsymbol{B}\times(\nabla\times\boldsymbol{M}))d\xi=$$

$$-\int_\xi\boldsymbol{r}\times(\nabla(\boldsymbol{B}\cdot\boldsymbol{M})-(\boldsymbol{B}\cdot\nabla)\boldsymbol{M}-(\boldsymbol{M}\cdot\nabla)\boldsymbol{B}-\boldsymbol{M}\times(\nabla\times\boldsymbol{B}))d\xi=\int_\xi[\boldsymbol{r}\times(\boldsymbol{B}\cdot\nabla)\boldsymbol{M}]d\xi\,. \tag{8}$$

Here we have taken into account that the integral $\int_\xi[\boldsymbol{r}\times\nabla(\boldsymbol{B}\cdot\boldsymbol{M})]d\xi$ can be transformed to the surface integral, where the magnetization $\boldsymbol{M}$ is vanishing; also we have taken into account that in the adopted approximation (***B***≈constant), the terms $(\boldsymbol{M}\cdot\nabla)\boldsymbol{B}$ and $\boldsymbol{M}\times(\nabla\times\boldsymbol{B})$ are vanishing, too. Integrating by parts the remaining integral in eq. (8), we obtain:

$$\int_\xi[\boldsymbol{r}\times(\boldsymbol{B}\cdot\nabla)\boldsymbol{M}]d\xi=\boldsymbol{\mu}\times\boldsymbol{B}\,, \tag{9}$$

We again prove this equality in components. For example, for the $z$-component we have:

$$M_{2z}=\int_\xi\left[x\left(B_x\frac{\partial M_y}{\partial x}+B_y\frac{\partial M_y}{\partial y}+B_z\frac{\partial M_y}{\partial z}\right)-y\left(B_x\frac{\partial M_x}{\partial x}+B_y\frac{\partial M_x}{\partial y}+B_z\frac{\partial M_x}{\partial z}\right)\right]d\xi=$$

$$\int_\xi\left[\left(B_x\frac{\partial(xM_y)}{\partial x}-B_xM_y+B_y\frac{\partial(xM_y)}{\partial y}+B_z\frac{\partial(xM_y)}{\partial z}\right)-\left(B_x\frac{\partial(yM_x)}{\partial x}+B_y\frac{\partial(yM_x)}{\partial y}-B_yM_x+B_z\frac{\partial(yM_x)}{\partial z}\right)\right]d\xi=$$

$$\mu_xB_y-\mu_yB_x=(\boldsymbol{\mu}\times\boldsymbol{B})_z$$

Further we evaluate the term responsible for the interaction of magnetization currents of a dipole with a magnetic field (third term in the integrand of eq. (6)). Here we notice that due to the adopted constancy of proper electric dipole moment, we get $\partial\boldsymbol{P}/\partial t=0$ in the rest frame of a dipole. However, for a moving dipole the stationary distribution of its charges yields $d\boldsymbol{P}/dt=0$, and hence $\partial\boldsymbol{P}/\partial t=-(\boldsymbol{V}\cdot\nabla)\boldsymbol{P}$. With the latter equality we derive:

$$-\int_\xi\boldsymbol{r}\times((\boldsymbol{V}\cdot\nabla)\boldsymbol{P}\times\boldsymbol{B})d\xi=-\int_\xi\boldsymbol{r}\times((\boldsymbol{V}\cdot\nabla)(\boldsymbol{P}\times\boldsymbol{B}))d\xi=\boldsymbol{V}\times(\boldsymbol{p}\times\boldsymbol{B}). \tag{10}$$

(Here we used the equality $(\boldsymbol{V}\cdot\nabla)\boldsymbol{P}\times\boldsymbol{B}=(\boldsymbol{V}\cdot\nabla)(\boldsymbol{P}\times\boldsymbol{B})$, which reflect the adopted constancy of B at the location of point-like dipole). For example, let us demonstrate the validity of eq. (10) for the $y$-component:

$$\int_\xi[z((\boldsymbol{V}\cdot\nabla)(\boldsymbol{P}\times\boldsymbol{B})_x)-x((\boldsymbol{V}\cdot\nabla)(\boldsymbol{P}\times\boldsymbol{B})_z)]d\xi=$$

$$\int_\xi\left[z\left(V_x\frac{\partial(\boldsymbol{P}\times\boldsymbol{B})_x}{\partial x}+V_y\frac{\partial(\boldsymbol{P}\times\boldsymbol{B})_x}{\partial y}+V_z\frac{\partial(\boldsymbol{P}\times\boldsymbol{B})_x}{\partial z}\right)-x\left(V_x\frac{\partial(\boldsymbol{P}\times\boldsymbol{B})_z}{\partial x}+V_y\frac{\partial(\boldsymbol{P}\times\boldsymbol{B})_z}{\partial y}+V_z\frac{\partial(\boldsymbol{P}\times\boldsymbol{B})_z}{\partial z}\right)\right]d\xi=$$

$$\int_\xi \left( V_x \frac{\partial [z(\boldsymbol{P}\times\boldsymbol{B})_x]}{\partial x} + V_y \frac{\partial [z(\boldsymbol{P}\times\boldsymbol{B})_x]}{\partial y} + V_z \frac{\partial [z(\boldsymbol{P}\times\boldsymbol{B})_x]}{\partial z} - V_z(\boldsymbol{P}\times\boldsymbol{B})_x \right) d\xi -$$

$$\int_\xi \left( V_x \frac{\partial [x(\boldsymbol{P}\times\boldsymbol{B})_z]}{\partial x} - V_x(\boldsymbol{P}\times\boldsymbol{B})_z + V_y \frac{\partial (x(\boldsymbol{P}\times\boldsymbol{B})_z)}{\partial y} + V_z \frac{\partial (x(\boldsymbol{P}\times\boldsymbol{B})_z)}{\partial z} \right) d\xi =$$

$$\int_\xi [(V_x(\boldsymbol{P}\times\boldsymbol{B})_z - V_z(\boldsymbol{P}\times\boldsymbol{B})_x)] d\xi = -[\boldsymbol{V}\times(\boldsymbol{p}\times\boldsymbol{B})]_y .$$

Here we again have taken into account that the volume integrals, where the functions of spatial coordinates are subjected to differentiation, can be transformed to the surface integrals, where polarization $\boldsymbol{P}$ is vanishing.

Finally, addressing to the last term of integrand of eq. (6) (the hidden momentum contribution), and taking into account that for stationary magnetization $\frac{\partial}{\partial t}(\boldsymbol{M}\times\boldsymbol{E}) = -(\boldsymbol{V}\cdot\nabla)(\boldsymbol{M}\times\boldsymbol{E})$, we obtain by analogy with eq. (10):

$$-\varepsilon_0 \int_\xi \left( \boldsymbol{r}\times\frac{\partial}{\partial t}(\boldsymbol{M}\times\boldsymbol{E}) \right) d\xi = \varepsilon_0 \int_\xi (\boldsymbol{r}\times(\boldsymbol{V}\cdot\nabla)(\boldsymbol{M}\times\boldsymbol{E})) d\xi = -\varepsilon_0 \boldsymbol{V}\times(\boldsymbol{\mu}\times\boldsymbol{E}). \quad (11)$$

Substituting eqs. (7), (9)-(11) into eq. (6), we obtain the expression for total torque exerted on a point-like dipole by an electromagnetic field:

$$\boldsymbol{T}_{total} = \boldsymbol{p}\times\boldsymbol{E} + \boldsymbol{\mu}\times\boldsymbol{B} + \boldsymbol{V}\times(\boldsymbol{p}\times\boldsymbol{B}) - \varepsilon_0 \boldsymbol{V}\times(\boldsymbol{\mu}\times\boldsymbol{E}), \quad (12)$$

where all quantities are evaluated in a laboratory frame.

Now applying eq. (12) to the problem of ref. [1] about the interaction of point-like charge and magnetic dipole, we obtain in the laboratory frame K:

$$\boldsymbol{T}_{total} = \boldsymbol{p}\times\boldsymbol{E} - \varepsilon_0 \boldsymbol{V}\times(\boldsymbol{\mu}\times\boldsymbol{E}), \quad (13)$$

where $\boldsymbol{p} = \varepsilon_0(\boldsymbol{V}\times\boldsymbol{\mu})$ is the relativistic polarization of the moving magnetic dipole. Further on, the term $\boldsymbol{p}\times\boldsymbol{E}$ can be presented in the form

$$\boldsymbol{p}\times\boldsymbol{E} = \varepsilon_0(\boldsymbol{V}\times\boldsymbol{\mu})\times\boldsymbol{E} = -\varepsilon_0\boldsymbol{E}\times(\boldsymbol{V}\times\boldsymbol{\mu}) = \varepsilon_0\boldsymbol{\mu}\times(\boldsymbol{E}\times\boldsymbol{V}) + \varepsilon_0\boldsymbol{V}\times(\boldsymbol{E}\times\boldsymbol{\mu}) = \varepsilon_0\boldsymbol{V}\times(\boldsymbol{E}\times\boldsymbol{\mu}), \quad (14)$$

where we have used the Jacobi identity and taken into account that for the problem in question the vectors $\boldsymbol{V}$, $\boldsymbol{E}$ are parallel to each other at the location of point-like dipole, and $\boldsymbol{E}\times\boldsymbol{V} = 0$. Hence, substituting eq. (14) into eq. (13), we get $\boldsymbol{T}_{total} = 0$ in the laboratory frame K, as it should according to the relativistic requirements.

We omit here the straightforward calculations, which show that the expression (12) for the torque on a point-like dipole with the constant proper electric and magnetic dipole moments can be derived with the torque expression (7) of ref. [1] based on Einstein-Laub formula (6) of [1].

Thus, in contrast to the claim made in ref. [1], we conclude that the approach based on the Lorentz force law complemented by hidden momentum contribution is fully compatible with the relativistic requirements and momentum conservation law. We provided an expression for the force density due to variation of hidden momentum (4) to be added to Lorentz force density in problems on interaction of magnetic dipoles (or magnetic media) with electromagnetic field. In particular, we have shown that a torque exerted on a magnetic dipole due to existence of its hidden momentum in an external electric field, defined by eq. (11), can be a non-zero value, even if the force (2) due to variation with time of hidden momentum is vanishing. The latter result, being applied to the interaction of point-like charge and magnetic dipole, as proposed in ref. [1], yields a consistent relativistic solution within the framework of the Lorentz force approach.